\documentclass[fleqn,10pt]{wlscirep}
\title{Inter-layer synchronization in non-identical multi-layer networks}

\author[1,2*]{I. Leyva}
\author[3]{R. Sevilla-Escoboza}
\author[1,2]{I. Sendi\~na-Nadal}
\author[4]{R. Guti\'errez}
\author[1,2]{J.M. Buld\'u}
\author[5,6]{S. Boccaletti}

\affil[1]{Complex Systems Group {\& GISC}, Universidad  Rey Juan Carlos, 28933 M\'ostoles, Madrid, Spain}
\affil[2]{Center for Biomedical Technology, Universidad Polit\'ecnica de Madrid, 28223 Pozuelo de Alarc\'on, Madrid, Spain}
\affil[3]{Centro Universitario de los Lagos, Universidad de Guadalajara, Jalisco 47460, Mexico}
\affil[4]{School of Physics and Astronomy, University of Nottingham, Nottingham, NG7 2RD, UK}
\affil[5]{CNR-Institute of complex systems, Via Madonna del Piano 10, 50019 Sesto Fiorentino, Italy}
\affil[6]{The Italian Embassy in Israel, Hamered Street 25, 68125 Tel Aviv, Israel}

\affil[*]{inmaculada.leyva@gmail.com}

\begin{abstract}
Inter-layer synchronization is a dynamical state occurring in multi-layer networks composed of identical nodes.
The state corresponds to have all layers synchronized, with nodes in each layer which do not necessarily evolve in unison.
So far, the study of such a solution has been restricted to the case in which all layers had an identical connectivity structure.
When layers are not identical, the inter-layer synchronous state is no longer a stable solution of the system.
Nevertheless, when layers differ in just a few links, an approximate treatment is still feasible, and allows one  to gather information
on whether and how the system may wander around an inter-layer synchronous configuration. We report the details of an
approximate analytical treatment for a two-layer multiplex, which results in the introduction of an extra inertial term
accounting for structural differences. Numerical validation of the predictions highlights the usefulness of our approach,
especially for small or moderate topological differences in the intra-layer coupling. Moreover, we identify a non-trivial relationship between the betweenness centrality of the missing links and the intra-layer coupling strength. Finally, by the use of two multiplexed identical layers of electronic circuits in a chaotic regime, we  study the loss of  inter-layer synchronization as a function of the betweenness centrality of the removed links.
\end{abstract}
\begin{document}

\flushbottom
\maketitle
\thispagestyle{empty}

\section*{Introduction}

Complex networks is one of the most active research topics in today's nonlinear science \cite{Boccaletti2006}.
As the field is rapidly evolving (mostly due to the huge amount of data collected nowadays), novel features are incorporated to better
describe real world systems.  Among these, the extension of the traditional framework to include  the {\it multi-layer} nature of networks has significantly altered the landscape of network science. In a multilayered description, units can be arranged in several layers (each of them accounting for a different kind of relationship or interaction between the nodes),
either simultaneously or alternatively  \cite{DeDomenico2013,kivela2014, Boccaletti2014}.

On the other hand, synchronization is one of the most relevant dynamical processes encountered in nature, and probably is the one that has been most thoroughly  studied in the context of complex networks \cite{Boccaletti2002,Boccaletti2006}. Only very recently the study of
synchronization has been extended to multi-layers \cite{Boccaletti2014} and, though an exact analytical treatment is available for just particular cases \cite{sorrentino2012,irving2012,Bogojeska2013,Aguirre2014}, several synchronization scenarios have been addressed. Namely, unidirectional coordination between layers \cite{Gutierrez2012, Lu2014}, multi-layer explosive synchronization \cite{Zhang2015}, synchronization driven by energy transport in interconnected networks \cite{Nicosia2014}, delayed synchronization between layers \cite{Louzada2013,Singh2015a} and global synchronization on interconnected layers as in Smart Grids \cite{Bogojeska2013} or neural systems \cite{Baptista2016}.
In the majority of these studies, the multi-layer structure of connections supports a global synchronous state
in which all the nodes in all the layers behave coherently.
More general forms of synchronization, however, are inherently possible on top of a multi-layer structure, as for instance  intra-layer synchronization \cite{Gambuzza2014} (where nodes evolve synchronously within each layer but layers do not necessarily evolve coherently), inter-layer  synchronization \cite{Gutierrez2012,Sevilla2016} (where, instead, layers are synchronized but nodes within each layer are not), and cluster synchronization  \cite{Jalan2016}.

Recently, we have provided analytical, numerical and experimental evidence of {\it inter-layer} synchronization \cite{Sevilla2016}, based on the assumption that different layers are topologically identical.
Here, we relax this assumption, and extend the study to the (much more realistic) case of nonidentical layers.
In particular, we offer a comprehensive (numerical, experimental and analytical) description  of the perturbative effects that the deletion of $m$ links  in one of the layers have on the  stability of the inter-layer synchronous state, and show the non-trivial relationship between the betweenness centrality of the missing links and the intra-layer coupling strength.

\begin{figure}[ht]
\centering \includegraphics[width=0.5\linewidth]{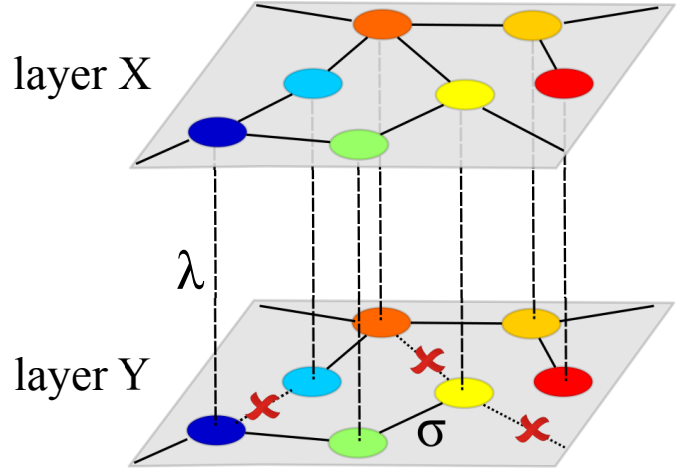}
 \caption{Schematic representation of a multiplex of two layers of identical oscillators. Labels $\sigma$ and $\lambda$ denote the intra- and inter-layer coupling strengths, respectively. Each node $i$ ($j$) in the top (bottom) layer is an $m$ dimensional dynamical system
 whose state is represented by the vector ${\bf x}_i$ (${\bf y}_j$).  The topologies of layers $X$ and $Y$ are encoded in the  $\mathcal{L}^1$
and $\mathcal{L}^2$ Laplacian matrices, respectively.  If we depart from two identical layers, we can write $\mathcal{L}^1 = \mathcal{L}^2 + \Delta \mathcal{L}$ where $\Delta \mathcal{L}$ contains the links that have been deleted in the bottom layer. }
\label{fig:net_noiden}
\end{figure}

\section*{Results}

The object of our study is a multiplex of two layers, formed by $N$ identical $m$ dimensional dynamical systems, whose states are represented by the vectors ${\bf X}=[{\bf x}_1,{\bf x}_2,\ldots,{\bf x}_N]^T$  and ${\bf Y}=[{\bf y}_1,{\bf y}_2,\ldots,{\bf y}_N]^T$
 with ${\bf x}_i, {\bf y}_i, \in \mathbb{R}^m$ for $i = 1,2,\ldots,N$.
Here, we focus on the case in which the topology of the two layers is different, and encoded by the elements of the Laplacian matrices $\mathcal{L}^{1}$ and $\mathcal{L}^2$ respectively, as depicted in Fig. \ref{fig:net_noiden}. Therefore, the evolution of the system is given by

\begin{eqnarray}\nonumber
  \label{eq:multiplex}
  \dot {\bf X}= {\bf F}({\bf X}) - \sigma \mathcal{L}^1 \otimes {\bf G}({\bf X}) + \lambda \,[{\bf H}({\bf Y})-{\bf H}({\bf X})],\\
  \dot {\bf Y}= {\bf F}({\bf Y}) - \sigma \mathcal{L}^2 \otimes {\bf G}({\bf Y}) + \lambda \,[{\bf H}({\bf X})-{\bf H}({\bf Y})],
\end{eqnarray}
where the functions ${\bf F}({\bf X})=[{\bf f}({\bf x}_1),{\bf f}({\bf x}_2),\ldots,{\bf f}({\bf x}_N)]^T$, ${\bf G}({\bf X})=[{\bf g}({\bf x}_1),{\bf g}({\bf x}_2),\ldots,{\bf g}({\bf x}_N)]^T$, and ${\bf H}({\bf X})=[{\bf h}({\bf x}_1),{\bf h}({\bf x}_2),\ldots,{\bf h}({\bf x}_N)]^T$, and ${\bf f}:\mathbb{R}^{m}\! \to\!\mathbb{R}^{m}$, ${\bf g}:\mathbb{R}^m\! \to\!\mathbb{R}^m$ and  ${\bf h}$: $\mathbb{R}^{m}\! \to\!\mathbb{R}^{m}$ represent, respectively, the local evolution (${\bf f}$) and the output vectorial functions within (${\bf g}$) and between (${\bf h}$) the layers. Parameters $\sigma$ and $\lambda$ are the intra- and the inter-layer coupling strengths.

When the layers are identical ($\mathcal{L}^1 = \mathcal{L}^2$), the inter-layer synchronous evolution (${\bf X}={\bf Y}$) is a solution of Eqs.~(\ref{eq:multiplex}), independently of the existence of intra-layer synchronization \cite{Sevilla2016}. When the inner structure of the layers differs ($\mathcal{L}^1 \neq \mathcal{L}^2$), however, ${\bf X}={\bf Y}$ is no longer a solution of Eqs.~(\ref{eq:multiplex}).
Yet, it can be expected that when the topologies of the two layers are actually {\it similar} (i.e. when their difference is limited to only a few links), one can proceed with an approximation, which consists in supposing that the dynamics of the system would anyway visit regions of the state space sufficiently close to ${\bf X}={\bf Y}$, so that the predictive use of the Master Stability Function (MSF) methodology
\cite{Pecora1998,Boccaletti2006} still makes sense.  In the Methods section, the interested reader can find the details of such an approximate MSF approach, whose predictions are tested in the following, both numerically and experimentally. It is, in any case, important to remark that our approach relies on approximations that are not fully controllable, and therefore one has to expect that predictions on the associated conditional Lyapunov exponents would less and less quantitatively fit the real evolution of the system, the more the two layers differ in the structure of connectivity.
The validity of the approximation is therefore checked by means of monitoring the value of the inter-layer synchronization error, which is defined as $E_{inter}=\lim_{T\to \infty}\frac{1}{T}\int_0^T\left\lVert \delta {\bf X}(t)\right\rVert dt$,  where $\delta {\bf X}(t)={\bf Y}(t)-{\bf X}(t)$ is the vector describing the difference between the layers' dynamics and $\lVert \rVert$ stands for the Euclidean norm. 

\subsection{Numerical results}

The first goal is to numerically assess the range of validity of the approximation.  For this purpose, the two layers are initially created as identical, and then structural differences are generated by removing $m$ links in $\mathcal{L}^2$. In order to evaluate the range of impact of the structural differences, we have chosen the $m$ links to be removed following an {\it edge betweenness} criterion \cite{Newman2004}. Accordingly, each simulation is repeated twice,  a first time removing the links with the highest edge betweenness ($m_+$), and a second time removing those that have the lowest edge betweenness  ($m_-$).  The procedure never produces a lack of connectedness in the graphs (for the networks and number of removals considered), and in case of  degeneracy, a link is chosen at random among those that have the same betweenness.  Without lack of generality, we consider two possible kinds of topologies where both layers
are either Erd\"os-R\'enyi \cite{erdos1959} (ER) or scale-free \cite{Barabasi1999} (SF), in all cases with $N=500$, and average degree $\langle k \rangle=8$.
\begin{figure}[ht]
\centering\includegraphics[width=0.8\linewidth]{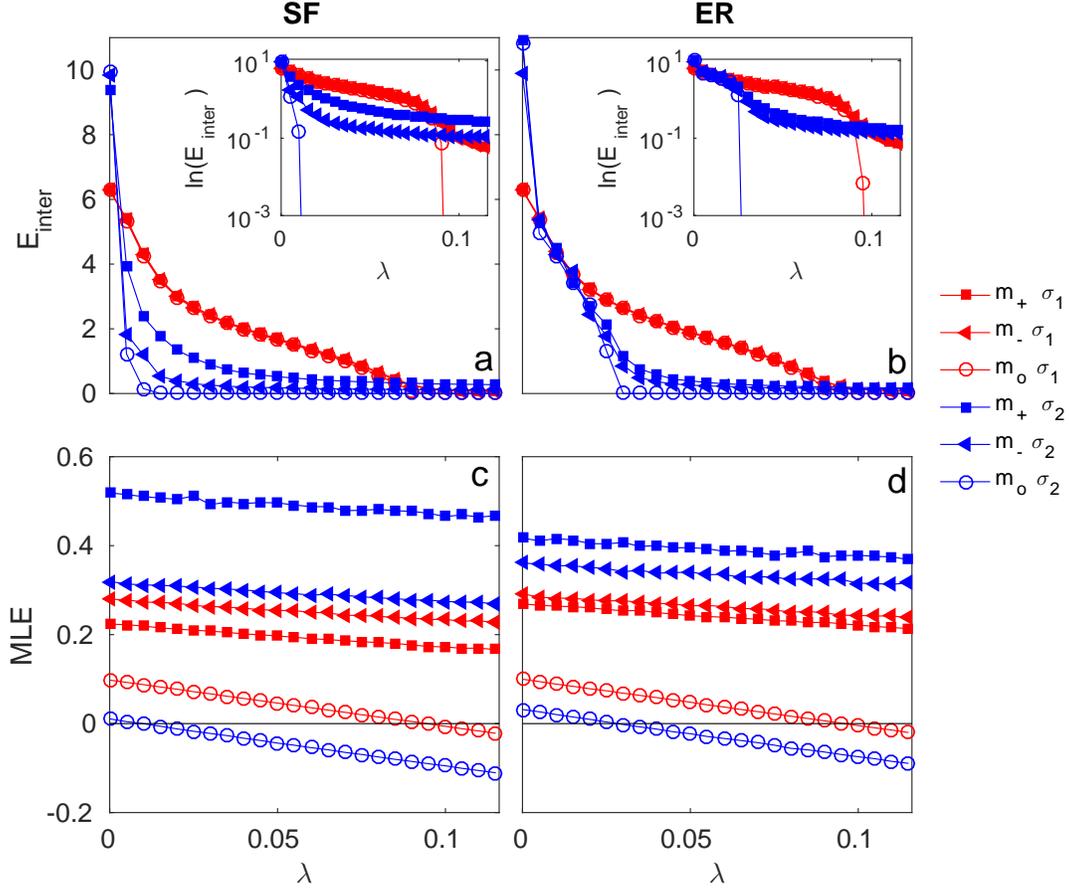}
 \caption{Results for inter-layer dynamics as a function of the intra-layer coupling strength $\lambda$ for class-I layers.
$E_{inter}$ (see main text) in multiplexes of SF (a) and ER (b) layers of $N=500$ R\"ossler oscillators, for two different values of intra-layer coupling $\sigma_1=0.1$ (red symbols) and $\sigma_2=1.0$ (blue symbols) when the 50 links with larger ($m_+$, $\blacksquare$) and lower ($m_-$, $ \blacktriangle$) betweenness are removed from $\mathcal{L}^2$, and for identical layers ($m=0$, $\circ$).
 Insets: Detail of the respective panels (a) and (b), in semi-logarithmic scale. (c)-(d) The corresponding MLE for the 
approximate expression in Eq.~(\ref{eq:variational_nonid}).}
\label{fig:zy_all}
\end{figure}

Nodes are here R\"ossler oscillators \cite{Rossler1976}, whose autonomous evolution is given by ${\bf f}({\bf x})=\left[-y-z,x+0.2 y,0.2+z(x-9.0)\right]$. ER and SF networks are generated by means of the procedures proposed in Refs.~\cite{erdos1959} and \cite{Barabasi1999}, respectively, and therefore the considered SF networks display a degree distribution $p(k) \propto k^{-3}$.

In our first example, the intra- and inter- layer coupling functions are set to be  $\mathbf{g}({\bf x})=(0,0,z)$ and  $\mathbf{{h}}({\bf x})=(0,y,0)$, respectively, so that (according to the standard MSF classification established in Ref.~\cite{Boccaletti2006}) the intra-layer configuration is within class I (and, therefore,  intra-layer synchronization is  never possible), whereas the inter-layer configuration corresponds to class II  (i.e., synchronization may be stable when the coupling strength exceeds a certain threshold).
\begin{figure}[ht]
\centering \includegraphics[width=0.5\linewidth]{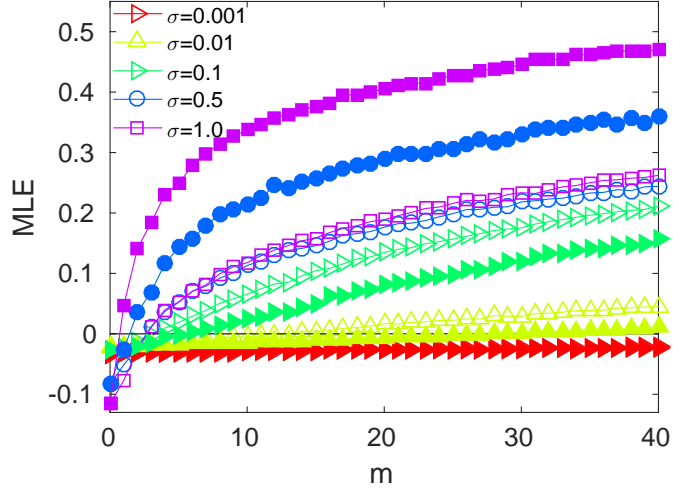}
 \caption{Maximum Lyapunov exponents (MLE) for different intra-coupling strengths $\sigma$ as a function of the number of links removed $m$, for the cases in which the removed links have the highest (full markers) or the lowest (void markers) edge betweenness. Layers are SF and of class I with $N=500$ R\"ossler oscillators and
$\lambda=0.12$.}
\label{fig:zy_mle_vs_m}
\end{figure}

In Fig.~\ref{fig:zy_all} we show the $E_{inter}$ (panels a and b) and MLE (panels c and d) as a function of the inter-layer coupling
$\lambda$ for two different values of intra-layer coupling $\sigma_1=0.1$ (red curves) and $\sigma_2=1.0$ (blue curves) when the 50 links
{(i.e. approximately $2.5\%$ of the total number)} with the largest ($m_+$, squares) and lowest ($m_-$, triangles)  betweenness centrality values are removed from the SF (Fig.~\ref{fig:zy_all}a,c) and ER (Fig.~\ref{fig:zy_all}b,d) $\mathcal{L}^2$ layers. For the sake of comparison, we also report
the curves for the case of identical layers ($m_0$, circles).

It can be observed that, in spite of the nonidentical layer topologies that make complete synchronization formally impossible,
the $E_{inter}$ series presents, in fact, apparently small differences with the identical case for both $m_+$ and $m_-$ and for the chosen $\sigma$ values, which can be better appreciated in a logarithmic representation (as shown in the insets of the corresponding figures).
Independently of the layer topology, at relatively large $\sigma$ ($\sigma_2$) the resilience of $E_{inter}$ to follow the trend observed in the identical case is larger than that corresponding to smaller values of $\sigma$ ($\sigma_1$).
This effect is in agreement with the fact that the non identicity of the layers results in the presence of an inertial term, which depends indeed on the value of $\sigma$ (see details in the Methods section). The corresponding Maximum Lyapunov Exponent (MLE) is shown in the bottom panels of Fig.~\ref{fig:zy_all}, confirming the behavior of the inter-layer dynamics depicted in the upper panels. Notice that the effects of removing links with high or low  betweenness are more pronounced in multiplexes made of SF layers
than in those made of ER ones. Another observation, which will be further highlighted in the following, is that the impact on the inter-layer
synchronization of removing high or low betweenness links is reversed depending on the strength of the intra-layer coupling: in both the ER and SF cases, removing $m_+$ links deteriorates (improves) the synchronization levels with respect to removing $m_-$ links for large (small) $\sigma$  values.

\begin{figure}[ht]
\centering  \includegraphics[width=0.6\linewidth]{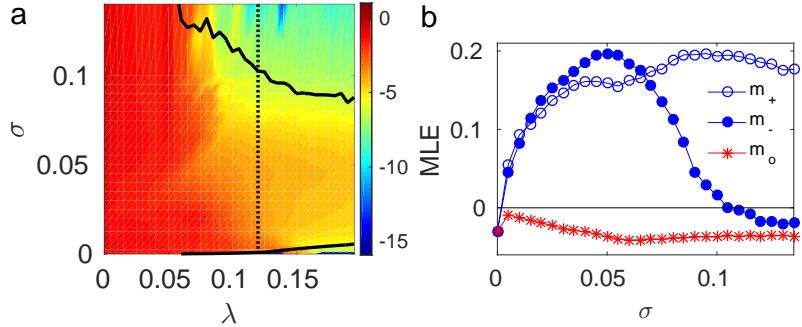}
 \caption{(a) Contour map of $\log(E_{inter})$ in the ($\sigma$, $\lambda$) parameter space with ${\bf g}({\bf x})= {\bf h}({\bf x}) = (x,0,0)$ (class III in the intra- and inter-layer dynamics) and $m_-=50$ links removed from  $\mathcal{L}^2$.  The black contour line corresponds to the isoline where the MLE changes its sign from positive to negative. Color code is shown in the lateral bar. (b) MLE vs. $\sigma$ for fixed 
$\lambda=0.12$ (corresponding to the dotted line in the left panel) where the 50 links with larger
($m=+50$, $\color{blue}\circ$) and lower ($m=-50$, $\color{blue}\bullet$) betweenness are removed from  $\mathcal{L}^2$. The identical case $m_0$ ($\color{red}*$)
is also plotted for comparison. In both panels, the two layers are ER of $N=500$ R\"ossler oscillators and $\langle k\rangle=8$.
Each point is an average over 5 realizations.}
\label{fig:numerical_xx}
\end{figure}

A better analysis of the role of the structural differences is provided in Fig.~\ref{fig:zy_mle_vs_m}, where  the dependence of MLE$(m_+)$ and MLE$(m_-)$ is reported as a function of $m$, for a fixed value of $\lambda$ (at which
there is inter-layer synchronization for $m=0$). As predicted by the approximated MSF approach, the dynamics drifts from the identical
case at smaller values of $m$,  as $\sigma$  increases. We here find a unexpected and interesting feature, already glimpsed before in Fig.~\ref{fig:zy_all}, that entangles the intra-layer
structure with the inter-layer dynamics: for small values of $\sigma$, removing the $m_-$ lowest betweenness centrality links
results in a stronger perturbation for the inter-layer synchronization than removing the same number $m_+$ of highest betweenness links.
However, for larger values of $\sigma$, the effect is reversed.

We tested also the case ${\bf g}({\bf x}) = {\bf h}({\bf x}) = (x,0,0)$ (where the MSFs belong to class III for both the intra- and inter-layer dynamics).
The validity of our approximation  is shown in the left panel of
Fig.~\ref{fig:numerical_xx}, where   $\log(E_{inter})$ is plotted for the $m_{-}=50$ case in the $(\sigma,\lambda)$ parameter space.
The limit in which the MLE becomes negative (black solid line) closely corresponds to $E_{inter}=0.001E_{0}$, being $E_0$ the corresponding inter-layer synchronization error for $\lambda=0$ (uncoupled layers) at each $\sigma$ value. Therefore, and once again, our approximated MSF provides an excellent reference for the analysis of the nonidentical inter-layer dynamics. In particular, in Fig.~\ref{fig:numerical_xx} we compare the MLE curves as a function of $\sigma$ in three different scenarios: identical layers (stars) and nonidentical layers after removal of the 50 links with lowest (full circles) and highest (empty circles) {betweenness}. In all cases, $\lambda$ was fixed to $0.12$ (which makes the two layers synchronizable when they are identical). For weakly coupled layers (low values of $\sigma$), the perturbation of removing $m_+$ or $m_-$ links is similar, but as the intra-layer coupling
increases, the multiplex is able to recover the inter-layer synchronization state despite {the} $m_-$ links {that} have been removed from one of the layers, while it is never again achieved in the case of removing the largest betweenness links.

\subsection{Experimental results}

\begin{figure}[ht]
\centering\includegraphics[width=0.7\linewidth]{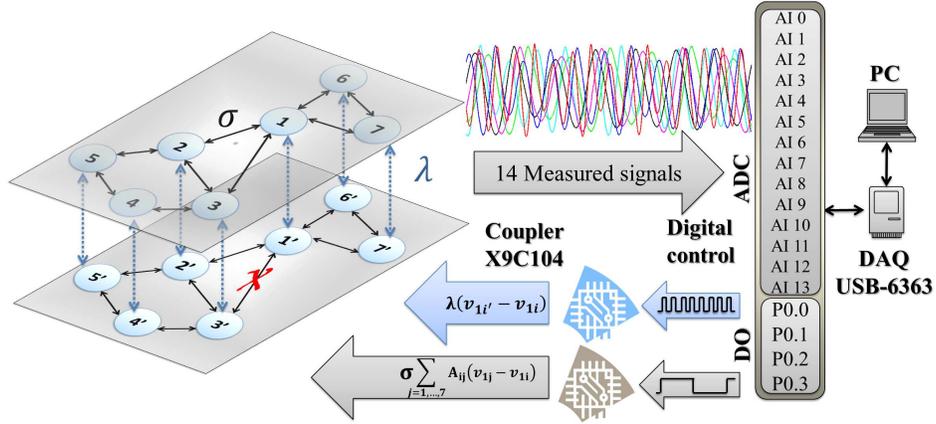}
  \caption{(Color online). Experimental setup. The left image is a sketch of the coupling topology of the 14 electronic circuits
composing the multiplex network
(see main text for the description of the experimental procedure used). The whole experiment is controlled from a PC with Labview Software.
\label{fig_setup}}
\end{figure}

Our predictions can be substantiated by an experiment with electronic circuits. The setup, sketched in Fig.~\ref{fig_setup}, consists of an electronic array, a personal computer (PC), 14 analog to digital converters (ADC) and 4 digital ports (DO) from a multi-functional data card (DAQ)
controlled by Labview. The ADCs are used for sampling one of the state variables out of all the networked circuits, {while} the DOs are used as controllers for the gain of the two coupling strengths $\sigma$ and $\lambda$. The array is made of 14 R\"ossler-like circuits arranged in two layers (blue nodes), each one of them having two different electronic couplers, one for the coupling among nodes in the same layer ($\sigma$) and the second for the interaction of each node with its replica in the other layer ($\lambda$). The layers are identical but for a single lacking link in one of the networks, which can be chosen to be any link in the experiment.

The chaotic dynamics of the {R\"ossler-like} circuits is well approximated by an electronic model, where  the nodes' state variables $x,y,z$ are translated into three different voltages $v_1$, $v_2$ and $v_3$:

\begin{eqnarray}
v_{1i}(t)&=&-\frac{1}{R_{1}C_{1}}\left(v_{1i} + \frac{R_{1}}{R_{2}}v_{2i}+\frac{R_{1}}{R_{4}}v_{3i} -\sigma\frac{R_{1}}{R_{15}} \sum_{j=1}^{N}{A_{ij} \left[ v_{1j}-v_{1i} \right] } -\lambda\frac{R_{1}}{R_{16}}\left[ v_{1^{'}i}-v_{1i} \right] \right) \label{eqrosslerExpx}\\
v_{2i}(t)&=&-\frac{1}{R_{6}C_{2}}\left(-\frac{R_{6}R_{8}}{R_{9}R_{7}}v_{1i}+ \left[1- \frac{R_{6}R_{8}}{R_{5}R_{7}}\right]v_{2i}  \right) \label{eqrosslerExpy}\\
v_{3i}(t)&=&-\frac{1}{R_{10}C_{3}}\left(-\frac{R_{10}}{R_{11}}G_{v_{1i}}+v_{3i} \right)
\label{eqrosslerExpz}
\end{eqnarray}
where $G_{v_{1i}}$ is a nonlinear gain function given by:
\begin{equation}
G_{v_{1i}}= \left\{ \begin{array}{lcc}
0 	&   \text{if}  & v_{1} \le Id+Id \frac{R_{14}}{R_{13}}+V_{ee}\frac{R_{14}}{R_{13}} \\
\\ \frac{R_{12}}{R_{14}}v_{1i}-V_{ee}\frac{R_{12}}{R_{13}}-Id\left(\frac{R_{12}}{R_{13}}+\frac{R_{12}}{R_{14}} \right)  &   \text{if}  & v_{1}> Id+Id \frac{R_{14}}{R_{13}}+V_{ee}\frac{R_{14}}{R_{13}} \\
\end{array}\right.
\label{eqrosslergx}
\end{equation}
and $C_i$ and $R_i$ are a series of capacitors and resistances whose values are summarized in Tab. \ref{TCexp} (the interested reader can have a look at Refs. \cite{Sevilla2015SR,Sevilla2016db} for a detailed description of the experimental implementation of the R\"ossler-like system, and  at Refs. \cite{Aguirre2014,Sevilla2015Le,Sevilla2016} for previous realizations with different network configurations).

\begin{table}[h] \centering
\begin{tabular}{|c|c|c|c|c|c|}
\hline \hline
$C_{1} = 4.7$nF			& $C_{2} = 4.7$nF	& $C_{3} = 4.7$nF 	&$\sigma = [0-0.25]$				\\
\hline \hline
$R1 = 2M\Omega$	& $R2 = 200K\Omega$ 	&$R3 = 10K\Omega$ 		& $R4 = 100K\Omega$ 		 \\
\hline
$R5 = 50K\Omega$	& $R6 = 5MK\Omega$ 	&$R7 = 100K\Omega$ 	& $R8 = 10K\Omega$ 		 \\
\hline
$R9 = 10K\Omega$	& $R10= 100K\Omega$ 	&$R11= 100K\Omega$ 	& $R12= 150K\Omega$ 		  \\
\hline
$R13= 68K\Omega$	& $R14= 10K\Omega$ 	&$R15= 100K\Omega$ 	&$R16= 100K\Omega$  		 \\
\hline
$RC= R3+R5$			&$Id = 0.7$  			&$V_{ee} = 15$ 		&$\lambda = [0-0.25]$  		 \\
\hline \hline
\end{tabular}
\caption{Values of the electronic components used for the construction of the electronic version of the R\"ossler-like system.}
\label{TCexp}
\end{table}

\begin{figure}[ht]
\centering\includegraphics[width=0.4\linewidth]{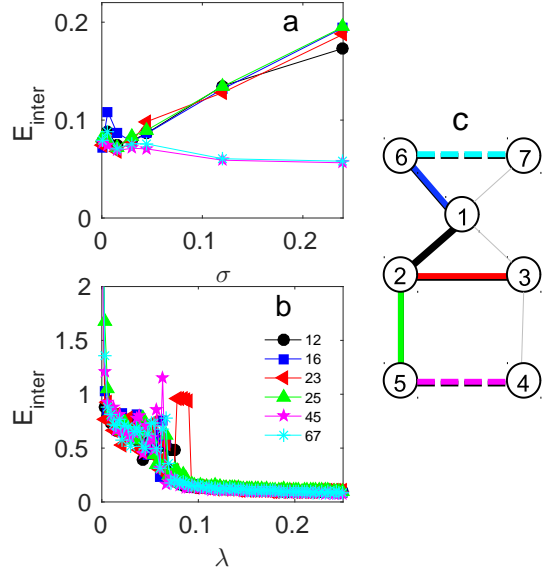}
\caption{Experimental results for a perturbed multiplex network of electronic R\"ossler oscillator. (a) $E_{inter}$ as a function of $\sigma$ for $\lambda=0.3$ and (b) $E_{inter}$ as a function of $\lambda$ for $\sigma=0.05$ for each case when one of the links (coded in legend of panel (b)) 
in one of the layers is removed. The color of the symbols corresponds to the color of the links 
in the layer structure scheme in panel (c). Data corresponding to high (low) edge betweenness links are drawn with solid (dotted) lines.  }
\label{fig:exper}
\end{figure}

{Departing from the initial network configuration of Fig.~\ref{fig:exper}c, we carry out a series of experiments where, one (different) link is removed from} one of the layers
{(always the same one)}. {The removed link between nodes $i$ and $j$} {will} be referred to in the following as $(ij)$. Both $\sigma$ and $\lambda$ values are initially set to zero, and the polarization voltage of the circuits is turned off and on, after a waiting time of 500 ms. The signals corresponding to the $x$ state variables of the 14 circuits are acquired by the analogue ports AI0-AI13 and saved in the PC for further analysis. For every $\sigma$ value, $\lambda$ is then incremented by one step, and the procedure is repeated 100 times (until the maximum value of $\lambda$ is reached). When the entire run is finished, $\sigma$ is increased by one step, and another cycle of $\lambda$ values is initiated. The whole procedure is repeated for every link of the network.

The experimental results for $E_{inter}$ are presented in Fig.~\ref{fig:exper}a,b, which confirm our predictions on the impact on the inter-layer dynamics of the removal of links with high or low betweenness. In Fig.~\ref{fig:exper}a,  as $\sigma$ {is increased while } keeping constant $\lambda=0.3$, the effect of deleting the links with the
highest {betweenness} [links (12),(16),(23) and (25) in our example, see network scheme in panel c for reference] {leads to a conspicuous increase in $E_{inter}$. A very different behavior is observed when we remove} the links with lower
{betweenness} [(45) and (67)], which consistently decreases the inter-layer error as the intra-layer coupling strength increases.
This is in full qualitative agreement with what observed  in the numerical counterpart (see Fig.~\ref{fig:numerical_xx}), and confirms the entanglement between the intra-layer structure and the inter-layer dynamics. Figure~\ref{fig:exper}b reports the dependence of  $E_{inter}$ on $\lambda$ when $\sigma$=0.05, showing that the network can reach a quasi-synchronous state even in the presence of structural defects, as predicted in Fig.~\ref{fig:numerical_xx}.
Also note the robustness of the theoretical predictions despite the intrinsic parameter mismatch  ($\sim$ 5\%) of the electronic components.

\section*{Discussion}

The obtained results allow us to draw a series of important conclusions about
the effects of structural layer differences on
the capability of multiplex networks to display synchronized layers, with nodes in each layer which do not necessarily evolve in unison.
It is important to remark that the study of inter-layer synchronization was restricted so far to the case in which all layers had an 
identical connectivity structure. When layers are not identical, several conceptual issues arise, the most relevant being that the inter-layer synchronous state is no longer a stable solution of the system, and one has therefore to proceed with approximate treatments.

We have demonstrated  that an approximate analytical treatment of a two-layer multiplex results in the introduction of an extra inertial term
accounting for structural differences. The predictions have been validated numerically and, most importantly, by means of an experiment with electronic circuits.
The conclusion is that, even in this case in which layers are not identical {and} the exact synchronized  solution does not exist, the approximate Master Stability Function is a very good
tool to study the inter-layer dynamics of {multiplex networks}. Using such a framework, indeed, we could predict the effect that missing links in one of the layer have on the inter-layer {synchronization}, evidencing a non-trivial relationship between the edge centrality of the different links and the balance between intra- and inter-layer couplings.

The fact that the predictions are solidly verified in an experimental setup (where fluctuations, noise and uncertainty of nodes' parameters are unavoidable)  highlights the robustness of our {analytical} predictions.

\section*{Methods}

\section*{Approximate Master Stability Function (MSF) formalism for a two-layer network}

We here summarize the main steps of the perturbation analysis of Eqs. \ref{eq:multiplex}.
First, one can always define $\delta {\bf X}(t) = {\bf Y}(t) - {\bf X}(t)=(\delta {\bf x}_1,\delta{\bf x}_2,\dots,\delta{\bf x}_N)$ and calculate its law of motion
\begin{eqnarray}\label{eq_delta}
\delta \dot{\bf X} &=& {{\bf F}}({\bf Y}) - {{\bf F}}({\bf X}) - \sigma  \mathcal{L}^2 \otimes {\bf G}({\bf Y}) + \sigma \mathcal{L}^1\otimes {\bf G}({\bf X})  - 2\lambda\,{{\bf H}(\delta{\bf X})} \nonumber
\end{eqnarray}
\noindent
where  ${\bf H}({\bf X})$ is assumed to be a linear function [i.e. ${\bf H}({\bf Y})-{\bf H}({\bf X}) = {\bf H}(\delta{\bf X})$]. Second, one can define $\Delta \mathcal{L} = \mathcal{L}^1 - \mathcal{L}^2$, as the matrix representing the difference between the two Laplacians. Plugging $\mathcal{L}^1 = \mathcal{L}^2 + \Delta \mathcal{L}$ into Eq.~(\ref{eq_delta}), one obtains the following dynamics at the level of individual nodes:

\begin{eqnarray}\label{eq_delta_nonid}
& \delta \dot{\bf x}_i &= {\bf f}({\bf y}_i) - {\bf f}({\bf x}_i) - 2 \lambda\, {\bf h}(\delta {\bf x}_i)
 - \sigma \sum_{k} \mathcal{L}^2_{ik}\, {\bf g}({\bf y}_k) + \sigma \sum_{k} \left(\mathcal{L}^2_{ik} + \Delta \mathcal{L}_{ik} \right) \, {\bf g}({\bf x}_k) \\
&  &= {\bf f}({\bf y}_i) - {\bf f}({\bf x}_i) - 2 \lambda\,  {\bf h}(\delta {\bf x}_i)   - \sigma \sum_{k} \mathcal{L}^2_{ik}\,\left[ {\bf g}({\bf y}_k) -  {\bf g}({\bf x}_k) \right] + \sigma \sum_{k} \Delta \mathcal{L}_{ik} \, {\bf g}({\bf x}_k) . \nonumber
\end{eqnarray}

Now, assume that in a large enough network the effect of the perturbation $\Delta \mathcal{L}$ is  small enough for an  inter-layer almost synchronous dynamics ${\bf y}_i(t) \approx {\bf x}_i(t)$ to emerge. Then, one can take $\delta {\bf x}_i$
to be small quantities, and expand to first order around ${\bf x}_i(t)$.
The equations resulting from the linearization are:

\begin{eqnarray}
\label{eq:variational_nonid}
 \delta \dot{\bf x}_i &=& \left[J{\bf f}({\bf \tilde x}_i) - 2 \lambda  J{\bf h}({\bf \tilde x}_i) \right] \delta {\bf x}_i - \sigma \sum_{k} \mathcal{L}^2_{ik}\, J{\bf g}({\bf \tilde x}_k) \delta {\bf x}_k   + \sigma \sum_{l} \Delta \mathcal{L}_{il} \, {\bf g}({\bf \tilde x}_l) \nonumber
\end{eqnarray}
 where ${\tilde{\bf x}}_i$ is the state of node $i$ in an isolated layer evolving according to ${{\dot{ \tilde {\bf x}}}}_{i} =  {\bf f}({{\tilde{\bf x}}}_i) - \sigma \sum_{k} \mathcal{L}^1_{ik}\, {\bf g}({ \tilde {\bf x}}_k)$.

 By comparing this result with the identical case \cite{Sevilla2016}, it can be seen that the non-identity of the systems  is reflected in the last {\it inertial} term, whose role in pushing the dynamics away from the identical case is expected to become more prominent when the topological differences  are large. Additionally, it predicts that the divergence from the inter-layer synchronization will depend on the  the intra-layer coupling strength, which is in its own right an interesting result on the rich interplay between intra-layer and inter-layer effects, an aspect of inter-layer synchronization that was thoroughly explored in the identical case in Ref~\cite{Sevilla2016}. Following the MSF approach, a negative sign in the maximum conditional Lyapunov exponent (MLE) obtained from Eqs.~(\ref{eq:variational_nonid}) can be taken as an indication for the presence of inter-layer synchronization \cite{Sevilla2016}.

%\bibliography{references}

\begin{thebibliography}{10}
\expandafter\ifx\csname url\endcsname\relax
  \def\url#1{\texttt{#1}}\fi
\expandafter\ifx\csname urlprefix\endcsname\relax\def\urlprefix{URL }\fi
\providecommand{\bibinfo}[2]{#2}
\providecommand{\eprint}[2][]{\url{#2}}

\bibitem{Boccaletti2006}
\bibinfo{author}{Boccaletti, S.}, \bibinfo{author}{Latora, V.},
  \bibinfo{author}{Moreno, Y.}, \bibinfo{author}{Chavez, M.} \&
  \bibinfo{author}{Hwang, D.-U.}
\newblock \bibinfo{title}{{Complex networks: Structure and dynamics}}.
\newblock \emph{\bibinfo{journal}{Phys. Rep.}} \textbf{\bibinfo{volume}{424}},
  \bibinfo{pages}{175--308} (\bibinfo{year}{2006}).

\bibitem{DeDomenico2013}
\bibinfo{author}{De~Domenico, M.} \emph{et~al.}
\newblock \bibinfo{title}{Mathematical formulation of multilayer networks}.
\newblock \emph{\bibinfo{journal}{Phys. Rev. X}} \textbf{\bibinfo{volume}{3}},
  \bibinfo{pages}{041022} (\bibinfo{year}{2013}).

\bibitem{kivela2014}
\bibinfo{author}{Kivel\"a, M.} \emph{et~al.}
\newblock \bibinfo{title}{Multilayer networks}.
\newblock \emph{\bibinfo{journal}{Journal of Complex Networks}}
  \textbf{\bibinfo{volume}{2}}, \bibinfo{pages}{203--271}
  (\bibinfo{year}{2014}).

\bibitem{Boccaletti2014}
\bibinfo{author}{Boccaletti, S.} \emph{et~al.}
\newblock \bibinfo{title}{{The structure and dynamics of multilayer networks}}.
\newblock \emph{\bibinfo{journal}{Phys. Rep.}} \textbf{\bibinfo{volume}{544}},
  \bibinfo{pages}{1--122} (\bibinfo{year}{2014}).

\bibitem{Boccaletti2002}
\bibinfo{author}{Boccaletti, S.}, \bibinfo{author}{Kurths, J.},
  \bibinfo{author}{Osipov, G.}, \bibinfo{author}{Valladares, D.} \&
  \bibinfo{author}{Zhou, C.}
\newblock \bibinfo{title}{{The synchronization of chaotic systems}}.
\newblock \emph{\bibinfo{journal}{Phys. Rep.}} \textbf{\bibinfo{volume}{366}},
  \bibinfo{pages}{1--101} (\bibinfo{year}{2002}).

\bibitem{sorrentino2012}
\bibinfo{author}{Sorrentino, F.}
\newblock \bibinfo{title}{{Synchronization of hypernetworks of coupled
  dynamical systems}}.
\newblock \emph{\bibinfo{journal}{New J. Phys.}} \textbf{\bibinfo{volume}{14}},
  \bibinfo{pages}{33035} (\bibinfo{year}{2012}).

\bibitem{irving2012}
\bibinfo{author}{Irving, D.} \& \bibinfo{author}{Sorrentino, F.}
\newblock \bibinfo{title}{Synchronization of dynamical hypernetworks:
  Dimensionality reduction through simultaneous block-diagonalization of
  matrices}.
\newblock \emph{\bibinfo{journal}{Phys. Rev. E}} \textbf{\bibinfo{volume}{86}},
  \bibinfo{pages}{056102} (\bibinfo{year}{2012}).

\bibitem{Bogojeska2013}
\bibinfo{author}{Bogojeska, A.}, \bibinfo{author}{Filiposka, S.},
  \bibinfo{author}{Mishkovski, I.} \& \bibinfo{author}{Kocarev, L.}
\newblock \bibinfo{title}{On opinion formation and synchronization in multiplex
  networks}.
\newblock In \emph{\bibinfo{booktitle}{Telecommunications Forum (TELFOR), 2013
  21st}}, \bibinfo{pages}{172--175} (\bibinfo{year}{2013}).

\bibitem{Aguirre2014}
\bibinfo{author}{Aguirre, J.}, \bibinfo{author}{Sevilla-Escoboza, R.},
  \bibinfo{author}{Guti\'{e}rrez, R.}, \bibinfo{author}{Papo, D.} \&
  \bibinfo{author}{Buld\'u, J.~M.}
\newblock \bibinfo{title}{{Synchronization of Interconnected Networks: The Role
  of Connector Nodes}}.
\newblock \emph{\bibinfo{journal}{Phys. Rev. Lett.}}
  \textbf{\bibinfo{volume}{112}}, \bibinfo{pages}{248701}
  (\bibinfo{year}{2014}).

\bibitem{Gutierrez2012}
\bibinfo{author}{Guti\'{e}rrez, R.}, \bibinfo{author}{Sendi{\~n}a-Nadal, I.},
  \bibinfo{author}{Zanin, M.}, \bibinfo{author}{Papo, D.} \&
  \bibinfo{author}{Boccaletti, S.}
\newblock \bibinfo{title}{{Targeting the dynamics of complex networks}}.
\newblock \emph{\bibinfo{journal}{Sci. Rep.}} \textbf{\bibinfo{volume}{2}},
  \bibinfo{pages}{396} (\bibinfo{year}{2012}).

\bibitem{Lu2014}
\bibinfo{author}{Lu, R.}, \bibinfo{author}{Yu, W.}, \bibinfo{author}{Lu, J.} \&
  \bibinfo{author}{Xue, A.}
\newblock \bibinfo{title}{{Synchronization on complex networks of networks}}.
\newblock \emph{\bibinfo{journal}{IEEE Transactions on Neural Networks and
  Learning Systems}} \textbf{\bibinfo{volume}{25}}, \bibinfo{pages}{2110--2118}
  (\bibinfo{year}{2014}).

\bibitem{Zhang2015}
\bibinfo{author}{Zhang, X.}, \bibinfo{author}{Boccaletti, S.},
  \bibinfo{author}{Guan, S.} \& \bibinfo{author}{Liu, Z.}
\newblock \bibinfo{title}{{Explosive Synchronization in Adaptive and Multilayer
  Networks}}.
\newblock \emph{\bibinfo{journal}{Phys. Rev. Lett.}}
  \textbf{\bibinfo{volume}{114}}, \bibinfo{pages}{038701}
  (\bibinfo{year}{2015}).

\bibitem{Nicosia2014}
\bibinfo{author}{Nicosia, V.}, \bibinfo{author}{Skardal, P.},
  \bibinfo{author}{Latora, V.} \& \bibinfo{author}{Arenas, A.}
\newblock \bibinfo{title}{{Spontaneous synchronization driven by energy
  transport in interconnected networks}}.
\newblock \emph{\bibinfo{journal}{arXiv:1405.5855v1}} \bibinfo{pages}{1--11}
  (\bibinfo{year}{2014}).

\bibitem{Louzada2013}
\bibinfo{author}{Louzada, V. H.~P.}, \bibinfo{author}{Ara\'{u}jo, N.},
  \bibinfo{author}{Andrade, J.~S.} \& \bibinfo{author}{Herrmann, H.~J.}
\newblock \bibinfo{title}{{Breathing synchronization in interconnected
  networks}}.
\newblock \emph{\bibinfo{journal}{Sci. Rep.}} \textbf{\bibinfo{volume}{3}},
  \bibinfo{pages}{3289} (\bibinfo{year}{2013}).

\bibitem{Singh2015a}
\bibinfo{author}{Singh, A.}, \bibinfo{author}{Ghosh, S.},
  \bibinfo{author}{Jalan, S.} \& \bibinfo{author}{Kurths, J.}
\newblock \bibinfo{title}{{Synchronization in delayed multiplex networks}}.
\newblock \emph{\bibinfo{journal}{EPL (Europhysics Letters)}}
  \textbf{\bibinfo{volume}{111}}, \bibinfo{pages}{30010}
  (\bibinfo{year}{2015}).
\newblock \eprint{1605.00352}.

\bibitem{Baptista2016}
\bibinfo{author}{Baptista, M.~S.}, \bibinfo{author}{Szmoski, R.~M.},
  \bibinfo{author}{Pereira, R.~F.} \& \bibinfo{author}{Pinto, S. E. D.~S.}
\newblock \bibinfo{title}{{Chaotic, informational and synchronous behaviour of
  multiplex networks.}}
\newblock \emph{\bibinfo{journal}{Sci. Rep.}} \textbf{\bibinfo{volume}{6}},
  \bibinfo{pages}{22617} (\bibinfo{year}{2016}).

\bibitem{Gambuzza2014}
\bibinfo{author}{Gambuzza, L.~V.}, \bibinfo{author}{Frasca, M.} \&
  \bibinfo{author}{G\'omez-Garde{\~n}es, J.}
\newblock \bibinfo{title}{Intra-layer synchronization in multiplex networks}.
\newblock \emph{\bibinfo{journal}{EPL (Europhysics Letters)}}
  \textbf{\bibinfo{volume}{110}}, \bibinfo{pages}{20010}
  (\bibinfo{year}{2015}).

\bibitem{Sevilla2016}
\bibinfo{author}{Sevilla-Escoboza, R.} \emph{et~al.}
\newblock \bibinfo{title}{{Inter-layer synchronization in multiplex networks of
  identical layers}}.
\newblock \emph{\bibinfo{journal}{Chaos: An Interdisciplinary Journal of
  Nonlinear Science}} \textbf{\bibinfo{volume}{26}}, \bibinfo{pages}{065304}
  (\bibinfo{year}{2016}).

\bibitem{Jalan2016}
\bibinfo{author}{Jalan, S.} \& \bibinfo{author}{Singh, A.}
\newblock \bibinfo{title}{Cluster synchronization in multiplex networks}.
\newblock \emph{\bibinfo{journal}{EPL (Europhysics Letters)}}
  \textbf{\bibinfo{volume}{113}}, \bibinfo{pages}{30002}
  (\bibinfo{year}{2016}).

\bibitem{Pecora1998}
\bibinfo{author}{Pecora, L.~M.} \& \bibinfo{author}{Carroll, T.~L.}
\newblock \bibinfo{title}{{Master Stability Functions for Synchronized Coupled
  Systems}}.
\newblock \emph{\bibinfo{journal}{Phys. Rev. Lett.}}
  \textbf{\bibinfo{volume}{10}}, \bibinfo{pages}{2109--2112}
  (\bibinfo{year}{1998}).

\bibitem{Newman2004}
\bibinfo{author}{Newman, M.~E.} \& \bibinfo{author}{Girvan, M.}
\newblock \bibinfo{title}{Finding and evaluating community structure in
  networks}.
\newblock \emph{\bibinfo{journal}{Phys. Rev. E}} \textbf{\bibinfo{volume}{69}},
  \bibinfo{pages}{026113} (\bibinfo{year}{2004}).

\bibitem{erdos1959}
\bibinfo{author}{Erd\"os, P.} \& \bibinfo{author}{R\'enyi, A.}
\newblock \bibinfo{title}{{On random graphs I.}}
\newblock \emph{\bibinfo{journal}{Publ. Math. Debrecen}}
  \textbf{\bibinfo{volume}{6}}, \bibinfo{pages}{290--297}
  (\bibinfo{year}{1959}).

\bibitem{Barabasi1999}
\bibinfo{author}{Barab\'asi, A.-L.} \& \bibinfo{author}{Albert, R.}
\newblock \bibinfo{title}{Emergence of scaling in random networks}.
\newblock \emph{\bibinfo{journal}{Science}} \textbf{\bibinfo{volume}{286}},
  \bibinfo{pages}{509--512} (\bibinfo{year}{1999}).

\bibitem{Rossler1976}
\bibinfo{author}{R\"{o}ssler, O.}
\newblock \bibinfo{title}{{An equation for continuous chaos}}.
\newblock \emph{\bibinfo{journal}{Phys. Lett.}} \textbf{\bibinfo{volume}{57}},
  \bibinfo{pages}{397--398} (\bibinfo{year}{1976}).

\bibitem{Sevilla2015SR}
\bibinfo{author}{Tirabassi, G.}, \bibinfo{author}{Sevilla-Escoboza, R.},
  \bibinfo{author}{Buld\'u, J.~M.} \& \bibinfo{author}{Masoller, C.}
\newblock \bibinfo{title}{Inferring the connectivity of coupled oscillators
  from time-series statistical similarity analysis}.
\newblock \emph{\bibinfo{journal}{Sci. Rep.}} \textbf{\bibinfo{volume}{5}},
  \bibinfo{pages}{10829} (\bibinfo{year}{2015}).

\bibitem{Sevilla2016db}
\bibinfo{author}{Sevilla-Escoboza, R.} \& \bibinfo{author}{Buld\'u, J.}
\newblock \bibinfo{title}{Synchronization of networks of chaotic oscillators:
  Structural and dynamical datasets}.
\newblock \emph{\bibinfo{journal}{Data in Brief}} \textbf{\bibinfo{volume}{7}},
  \bibinfo{pages}{1185--1189} (\bibinfo{year}{2016}).

\bibitem{Sevilla2015Le}
\bibinfo{author}{Sevilla-Escoboza, R.}, \bibinfo{author}{Buld\'u, J.~M.},
  \bibinfo{author}{Pisarchik, A.~N.}, \bibinfo{author}{Boccaletti, S.} \&
  \bibinfo{author}{Guti\'errez, R.}
\newblock \bibinfo{title}{Synchronization of intermittent behavior in ensembles
  of multistable dynamical systems}.
\newblock \emph{\bibinfo{journal}{Phys. Rev. E}} \textbf{\bibinfo{volume}{91}},
  \bibinfo{pages}{032902} (\bibinfo{year}{2015}).

\end{thebibliography}

\section*{Acknowledgments}

Work partly supported by the Spanish Ministry of Economy (under projects FIS2012-38949-C03-01 and FIS2013-41057-P), the Mexican University of Guadalajara, CULagos  (under projects PRO-SNI/228069, PROINPEPRG/005/2014 and
UDG-CONACyT/I010/163/2014), and  by GARECOM, Group of Research Excelence URJC-Banco de Santander.
Authors acknowledge the computational resources and assistance provided by CRESCO, the super-computing center of ENEA in Portici, Italy.

\section*{Author contributions statement}

IL, ISN, RG and SB conceived the study, devised the model network, and prepared figures. RS and JB carried out the experiments. IL and ISN carried out the numerical simulations. All Authors wrote the Manuscript.

\section*{Additional information}

\textbf{Competing financial interests}: There are no competing financial interests.

\end{document}